# Molecular interactions between the constituents of small ribosomal subunit

Saurav Mallik and Sudip Kundu[*]

Department of Biophysics, Molecular Biology and Bioinformatics, University of Calcutta; Address: 92, APC Road, Kolkata; India; Zip: 700009

**ABSTRACT**
Availability of high-resolution crystal structures of ribosomal subunits of different species opens a route to investigate about molecular interactions between its constituents and stabilization strategy. Structural analysis of the small ribosomal subunit shows that primary binder proteins are mainly employed in stabilizing the folded ribosomal RNA by their high negative free energy of association, where tertiary binders mainly help to stabilize protein-protein interfaces. Secondary binders perform both the functions. Conformational changes of prokaryotic and eukaryotic ribosomal proteins due to complexation with 16S ribosomal RNA are linearly correlated with their RNA-interface area and free energy of association. The proteins having long extensions buried within ribosomal RNA have more flexible structures than those found on the subunit surface. *Thermus thermophilus* ribosomal proteins undergo high conformational changes compared to those of *Escherichia coli*, assuring structural stability at high temperature environment. The general stabilization strategy of ribosomal protein-RNA interfaces is shown, where high interface polarity ensures high surface density of Hydrogen bonds even with low base/backbone ratio. Polarity is regulated in evolutionary strategy of ribosomal proteins. Thus, the habitat environmental conditions of the two species sweet up their ribosomal protein-RNA interfaces to alter its physical parameters in order of stabilization.

## 1 INTRODUCTION

Ribosomes are giant ribonucleoprotein complexes that act as polymerases translating the genetic code into protein by providing the framework of positioning of the other participants in this process (Yonath, 2009; Moore, 1998). The prokaryotic ribosome consists of a larger 50S subunit and a smaller 30S subunit, of which high resolution X-ray crystallographic structures have been resolved (Wimberly *et al.* 2000; Ban *et al.* 2000; Schluenzen *et al.* 2000; Brodersen *et al.* 2002, Ben-Shem *et al.*, 2010, Korostelev *et al.*, 2006, 2009; Schuwirth *et al.*, 2005) and computationally analyzed (Noller and Baucon, 2002; Tung and Sanbonmatsu, 2004; Taylor *et al.*, 2009). Due to the functional importance of the ribosome, it is interesting to investigate about the stability of the overall complex and the interactions among its various components. The 30S subunit consists of 16S ribosomal RNA (16S rRNA) and about 21 proteins (Agalarov *et al.* 2000; Klein *et al.* 2004). The structures and positions of the proteins on 16S rRNA have been described in detail by Brodersen *et al.* 2002, Agalarov *et al.* 2000 etc. Assembly pathway of the constituents of small ribosomal subunit was analyzed much earlier. The 30S subunit has shown to be reconstituted from purified components in vitro, and the ordered nature of the assembly was first revealed by the elegant work of Held *et al.* 1974, known as 'Nomura Model'. An advancement of this concept is done by Mulder *et al.*, 2010 and Chen *et al.* in 2012. According to the joining with 16S rRNA during the 30S assembly pathway, the smaller subunit proteins have been divided into three classes (Held *et al.* 1974): the primary binding proteins (binds to the naked 16S rRNA: S4, S7, S8, S15 and S20), the secondary binding proteins (requires binding sites located at the RNA and primary binders to associate: S9, S12, S13, S16, S18 and S19) and the tertiary binding proteins (requires binding sites located at the RNA, primary and secondary binders to be associated: S2, S3, S5, S6, S10, S11, S14, S21 and THX).

Availability of many high-resolution crystal structures in PDB (Bernstein *et al.* 1977) has opened the possibility to analyze the stabilization strategy of 30S and 50S subunits by mutual interaction between RNA and protein components of them. In addition, the availability of crystal structures of two different species (*Escherichia coli* and *Thermus thermophilus*) being habituated in two different habitat conditions offers an opportunity to investigate whether habitat conditions have any effect on stabilization strategy of ribosomal subunits. In the current work, we have presented a detailed analysis on the molecular interactions stabilizing the 30S subunit structure. We present the relative influence of ribosomal proteins (classified according to Nomura model) in stabilization of the folded 16S rRNA at various stages of 30S assembly. We have estimated the conformational changes of the ribosomal proteins due to complexation with 16S rRNA. The linear correlation of the conformational changes of proteins on their rRNA-interface area and free energy of association has been established. A comparison of ribosomal and non-ribosomal protein-RNA interface properties establishes the unique nature of the former compared to the later. The regulation of interface polarity generates high surface density of Hydrogen bonds at the protein-rRNA interfaces, which ensures its structural stability. Interface polarity is regulated in evolutionary strategy of ribosomal proteins according to habitat conditions.

## 2 MATERIALS AND METHODS

Details of the materials and methods are included in the supplementary materials.

---

[*]To whom correspondence should be addressed.

## 2.1 Dataset

The atomic coordinate of 21 small ribosomal subunits were extracted from the PDB archive (Bernstein *et al.* 1977) on 16/05/2012. All the structures containing modified residues, full-length mRNA and tRNA, elongation factors and release factors were neglected. Eleven of our selected structures are of *Thermus thermophilus* (prokaryote); nine are of *Escherichia coli* (prokaryote) and one of Tetrahymena *thermophila* (eukaryote). A detailed description of the dataset has been included in the Supplementary Table S1. Classification of small subunit proteins are enlisted in Supplementary Table S2.

## 2.2 Computational Analysis

Details of the computational analysis are included in Supplementary materials.

We identified the protein-RNA and protein-protein interfaces of the 30S ribosomal subunit and several interface properties were calculated. These properties are buried surface area (BSA), free energy of interface formation (ΔG), surface density of Hydrogen-bonding (H-bonding) and van der Waals contacts at those interfaces, interface polarity, base/backbone ratio of protein-RNA H-bonding, conformational changes of the ribosomal proteins due to their interaction with 16S rRNA. In addition, we have estimated the conservation at the interfaces of ribosomal proteins.

## 3 RESULTS AND DISCUSSION

### 3.1 Influence of ribosomal proteins in 30S subunit stabilization

#### 3.1.1 Protein-RNA interactions

We first analyze contributions of ribosomal proteins in stabilizing the final 30S structure through their interactions with 16S rRNA. The solvent accessible surface area of each protein buried against (BSA) the 16S rRNA was calculated using Surface racer program (Tsodikov *et al.*, 2002) and their free energies of association (ΔG) were calculated using PDBePISA server (Krissinel and Henrick, 2005, 2007; Krissinel 2009). The results are listed in Table-1 as average between two species (*Thermus thermophilus* and *Escherichia coli*). It is expected that the proteins having large BSAs against 16S rRNA and high negative ΔG play most important roles in stabilizing the folded 16S rRNA.

**Table 1:** The buried surface area (BSA) and free energy of association (ΔG) at protein-rRNA interface of ribosomal proteins are displayed. Proteins have been classified both according to Nomura model. Corresponding standard deviation values are presented in the parenthesis.

| Protein Groups | | Buried surface area (Å$^2$) | Free energy of association (ΔG) with 16S rRNA (Kcal/mole) |
|---|---|---|---|
| Classification according to Nomura model, 1974 | Primary | 2091.75 (525.74) | -36.47 (12.49) |
| | Secondary | 2337.80 (683.37) | -42.31 (16.37) |
| | Tertiary | 1696.13 (375.94) | -29.05 (7.85) |

The results show that the secondary binding proteins have the largest average BSAs and ΔGs and tertiary binders have the lowest. Primary binders associate with 16S rRNA with BSA and ΔG values slightly less than secondary binders. The average BSA of the primary binding proteins is 2091.75 Å$^2$ and that for secondary binders is 2337.80 Å$^2$. The free energies of association of the primary and secondary binders are -36.47 Kcal/mole and -42.31 Kcal/mole, respectively. The comparatively much lower average BSA (1696.13 Å$^2$) and ΔG (-29.05 Kcal/mole) of tertiary binding proteins with 16S rRNA indicate that these proteins are less important in stabilization of the folded 16S rRNA compared to primary and secondary binders. The BSA and ΔG values of the individual proteins for both *E. coli* and *T. thermophilus* have been enlisted in Supplementary Table S3 and S4, respectively.

The statistical significance test (Supplementary Table S5) shows that the primary and secondary binding proteins are originated from the same populations. On the other hand, the difference of this population from tertiary binders is marginally significant (p=0.03). In addition, we generate a hierarchical cluster (using UPGMA method: Gronau and Moran, 2007) of proteins using their BSAs. Four distinct groups (Supplementary Figure F1) were identified; and among them, two groups (Group-1 and Group-2) with high BSAs and high negative ΔG of associations (statistically significantly different from other groups) consist of three primary (S4, S17 and S20) and three secondary (S9, S12 and S16) binders proteins (Table S6). Thus, the primary and secondary binders play more important role in stabilizing protein-rRNA complexed structure, compared to tertiary binders.

### 3.1.2 Protein-protein interactions

Next, we study how the ribosomal proteins mutually interact to reduce the overall ΔG value of the 30S subunit structure and contribute further to its stabilization. The results are included in Table 2. The corresponding values for *T. thermophilus* and *E. coli* are presented in Supplementary Table S7.

**Table 2:** The protein-protein interface properties of *T. thermophilus* and *E. coli* ribosomes; abbreviations used in this table: σ(H)=number of H-bonds per 100Å$^2$ buried surface area (interface area), σ(v)= number of van der Waals contacts per 100Å$^2$ buried surface area, ΔG= free energy of interface formation.

| Protein-protein interactions classified according to association | | Interface Properties |
|---|---|---|
| Classification | Protein-protein interfaces | |
| Primary-Primary | S8-S17 | σ(H)= 2.60<br>σ(v)= 18.00<br>ΔG= -0.70 |
| Primary-Secondary | S7-S9, S8-S12, S12-S17 | σ(H)= 0.30<br>σ(v)= 1.93<br>ΔG= -2.10 |
| Primary-Tertiary | S2-S8, S3-S4,<br>S4-S5, S5-S8, S7-S11 | σ(H)= 1.17<br>σ(v)= 10.63<br>ΔG= -0.76 |
| Secondary-Secondary | S9-S10, S9-S13, S13-S19 | σ(H)= 0.95<br>σ(v)= 12.65<br>ΔG= -1.17 |
| Secondary-Tertiary | S6-S18, S9-S14, S11-S18, S14-S19,<br>S18-S21 | σ(H)= 10.63<br>σ(v)= 46.45<br>ΔG= -2.49 |
| Tertiary-Tertiary | S2-S3, S2-S5, S3-S5, S3-S10, S3-S14,<br>S10-S14, S11-S21 | σ(H)= 1.80<br>σ(v)= 14.68<br>ΔG= -4.23 |

An average secondary-tertiary interface is associated with -2.49 Kcal/mole free energy, where tertiary-tertiary interfaces are associated with average -3.08 Kcal/mole free energy. Thus, they form very stable interfaces among themselves. The secondary-secondary interfaces are also very stable (with average -1.17 Kcal/mole free energy of association). However, primary-primary and primary-tertiary interfaces have lower values of ΔG in comparison to the other interactions. Interestingly, while the tertiary binders have a lower ΔG of interface formation with primary binders (-0.76 Kcal/mole); the secondary binders have higher values of ΔG of interface formation with primary binders (-2.10 Kcal/mole). The average values of surface densities of van der Waals contacts of tertiary binders (secondary-tertiary, primary-tertiary and tertiary-tertiary) are higher than those of other interfaces. The tertiary binders do also have the highest surface density of hydrogen bonds with the primary binders.

### 3.1.3 The roles of proteins at various stages of 30S subunit protein-RNA assembly

In case of protein-RNA interactions of 30S subunit, we observe that the primary and secondary binding proteins have higher negative free energies of association with the 16S rRNA (Table 1). Table 2 depicts that the average solvent free energy changes due to protein-protein associations are remarkably high for tertiary-tertiary and secondary-tertiary interactions compared to the others. Thus, we find that primary binders significantly stabilize the protein-rRNA interfaces and tertiary binders mainly stabilize the protein-protein interfaces of 30S subunit. Secondary binders play both the roles significantly.

The above results clearly indicate the possible underlying biophysical principles during the ribosomal assembly. In the initial states of 30S assembly process, the proteins bind to the naked 16S rRNA, which has a very high folding free energy (isolated structure has about +4000 Kcal/mole solvation free energy for *E. coli*, calculated using PDBePISA server). Therefore, the unfolding of naked 16S rRNA in the initial stages of assembly is highly favorable. The primary binding proteins, which associate first with 16S rRNA, play an important role in stabilizing the folded 16S rRNA initially and prevent the unfolding. The secondary binders then associate and contribute further to the rRNA stabilization. In addition, they form stabilized protein-protein interfaces with primary and tertiary binders that cause the 30S ribosome structure intact. Finally, tertiary binders, with their high free energies of protein-protein associations and high inter surface densities of H-bonds and van der Waals contacts provide the final stability to the 30S ribosomal complex.

## 3.2 Conformational changes of 30S subunit proteins upon complexation with 16S rRNA

Here, we study the conformational changes of the ribosomal proteins due to their interaction with the 16S rRNA and its dependency on other physico-chemical properties.

### 3.2.1 Protein conformational changes

Most of the ribosomal proteins have several extended domains buried in RNA; being isolated from the subunit, they do not contain these extensions (Brodersen *et al.*, 2002). Thus, the usual structural alignment method (comparison of un-complexed and complexed form) is not a suitable one in this case to measure the conformational changes due to complexation. Therefore, we have used another algorithm, (Marsh and Teichmann, 2011), which can predict the probable conformational change occurring in a protein monomer due to complexation with its partner. (see Supplementary Materials, section 1.4). Conformational changes are expressed in terms of Root Mean Square Distance (RMSD) between the aligned uncomplexed and complexed structures. The predicted RMSD values between complexed and uncomplexed states of the ribosomal proteins are presented in Table 2. The predicted RMSD values vary in a wide range for both the species. Bellow we explain the significance of conformational changes of individual proteins.

**Table 3:** The predicted conformational changes of the ribosomal proteins due to their complexation within the 30S ribosomal subunit are listed here. Conformational changes are expressed in terms of Root Mean Square Distance (RMSD) between the aligned uncomplexed and complexed structures.

| Ribosomal proteins | Predicted RMSD between complexed and uncomplexed states (Å) | |
|---|---|---|
| | *T. thermophilus* | *E. coli* |
| S2 | 2.98 | 2.63 |
| S3 | 5.61 | 5.01 |
| S4 | 4.71 | 3.66 |
| S5 | 2.99 | 2.93 |
| S6 | 3.89 | 1.67 |
| S7 | 7.69 | 5.70 |
| S8 | 3.07 | 2.59 |
| S9 | 6.56 | 5.00 |
| S10 | 9.12 | 7.57 |
| S11 | 5.56 | 4.88 |
| S12 | 18.76 | 11.14 |
| S13 | 14.25 | 8.10 |
| S14 | 19.19 | 17.50 |
| S15 | 5.78 | 3.90 |
| S16 | 2.25 | 3.04 |
| S17 | 9.79 | 5.08 |
| S18 | 5.04 | 3.31 |
| S19 | 6.83 | 4.80 |
| S20 | 8.94 | 5.72 |
| S21 | - | 32.26 |
| THX | 4.30 | - |

Brodersen *et al.* (2002) showed that S2, S4, S5 and S8 proteins have large globular domains which are found on the surface of 30S subunit; they usually have small extended regions buried within 16S rRNA. On the other hand, S12, S13, S14, S17 and S19 have comparatively long rRNA buried extended regions. Authors hypothesized that proteins having long RNA-buried extensions assist 16S rRNA folding. Thus, it follows that such proteins should undergo high conformational changes during 16S rRNA folding and should have more flexible structures compared to the proteins found on the surface of 30S subunit.

Following the work of Marsh and Teichmann, we predict the flexibilities of small subunit proteins (Supplementary Table S8a) and clustered them (UPGMA method) into three groups in increasing order: (Group-A) S2, S5, S6, S8, S16; (Group-B) S3, S4, S7, S9, S11, S15, S19; (Group-C) S10, S17, S18, S20; (Group-D) S12, S13, S14. Group-A and B proteins are found on 16S rRNA surface. They have small RNA buried extensions, rigid structures and undergo small conformational changes upon interaction with 16S rRNA (average RMSD on both species are 2.80 and 5.14 respectively). Group-C proteins have intermediate structural flexibility, comparatively longer RNA buried extensions; they undergo intermediate structural changes (average RMSD=7.45). Group-D proteins have long RNA buried extensions and highly flexible structures which experience large structural changes (average RMSD=14.82). Difference between Group-A and B (p(A-B)=0.001) and Group-B and C (p(B-C)=0.004) are significant. Group-D is marginally different from every other group (p(C-D)=0.05; p(A-D)=0.035; p(B-D)=0.012). Group-A and C are also marginally different (p(A-C)=0.015).

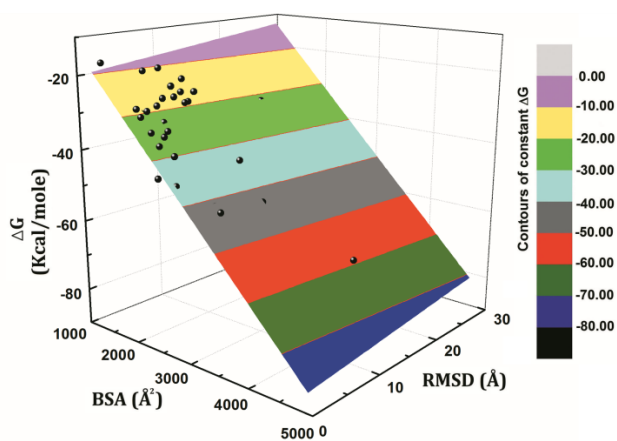

**Figure-1:** The 3D surface plot for ΔG, RMSD and BSA is shown here. The surface is presented as 3D colormap, with projection on the BSA-RMSD plane shows the contours of constant ΔG. Black dots represent the data points (not all are visible as some data points are beneath the surface). This plot is generated by Origin data analysis and graphic workspace (37), using the ribosomal protein-RNA association data from two bacteria (*E. coli* and *T. thermophilus*) and one eukaryotic species (*T. thermophila*).

It also appears from Marsh and Teichmann's work that S12 and S14 (for both *E. coli* and *T. thermophilus*) are likely to be intrinsically disordered, which also follows from predictions of DisEMBL server (Linding *et al.*, 2003). Marsh and Teichmann showed (for protein-protein interactions) that upon complex formation, intrinsically disordered protein monomers undergo high structural changes. This might entail the fact that disordered proteins often show disorder-to-order transitions upon complex formation (Wright, 1999; Fink, 2005; Dunker *et al.*, 2002; Radivojac *et al.*, 2004). The high conformational change of S12 and S14 during its assembly with 16S rRNA might be the result of such a disorder-to-order transition event. Another interesting fact is that, most of the proteins of *T. thermophilus* experience higher conformational changes compared to those of *E. coli*. Implication of this is discussed in section 3.3.

### 3.2.2 Protein conformational changes depend on BSA and ΔG of protein-rRNA interfaces

Next, we have tried to explore whether there exist any relationship between conformational changes, and the strength of protein-rRNA interaction, represented by ΔG of association and BSA. Earlier works (Janin *et al.*, 1988; Lo Conte *et al.*, 1999) have shown in case of protein-protein interactions, the conformational changes of the monomers depend on their interface area. However, in ribosomal protein-RNA interactions (using data for the three species together), we observe a very low value of linear correlation coefficient (r=0.35, R-square=0.13) between conformational change (RMSD) and BSA. We also observe a poor linear fitting of RMSD and ΔG of association (r=0.27, R-square=0.09). Interestingly, the three parameters, all-together, correlate very well with each other in a linear equation (r=0.93 and R-square=0.86), defining a plane surface in a three-dimensional space. The linear equation is shown in the following:

$$\Delta G = a + b(BSA) + c(RMSD) \qquad (1)$$

Here a, b and c are the fitting constants. This very equation is observed in both the bacterial species (*T. thermophilus* and *E. coli*) and is followed in an eukaryotic ribosome (*Tetrahymena thermophila*) as well. However, the values of the coefficients, a, b and c are species specific. In Supplementary Table S8b, the surface fitting statistics along with the values of these three constants for the three species have been included. The high correlation among the three parameters entails that the conformational changes of the ribosomal proteins are not solely dependent on the BSA or ΔG of association. In fact, both the physical parameters jointly determine the conformational changes of the ribosomal proteins.

The 3D surface plot using the average BSA, RMSD and ΔG values of the three species (*T. thermophilus, E. coli* and *T. thermophila*) is presented in Figure-1. It is follows from Equation-1 that proteins with high negative ΔG of association with 16S rRNA and large interface areas experience more conformational changes upon interaction compared to those having smaller interfaces and low ΔG of association. Nevertheless, when we modify our data taking into account protein-protein interactions, this linear correlation becomes very poor. This probably indicates the fact that major structural changes of the ribosomal proteins occur due to complexation with the 16S rRNA only.

### 3.3 Generalized Stabilization Strategy of 30S Subunits in both Bacterial Species

While studying the *E. coli* and T. *thermophilus* separately, we observe that the same strategy is followed by both the species to stabilize their 30S subunit assembly. However, there exists difference in the corresponding values of interface parameters of two species, which can be correlated with the habitat conditions they live in. Our results are discussed in the following.

**3.3.1 Primary and Secondary Binder play major roles in stabilizing 16S rRNA structure:** We list below the free energies of 16S rRNA-protein associations. The values for primary binders are -39.51 Kcal/mole and -35.84 Kcal/mole for *T. thermophilus* and *E. coli*, respectively. Those for secondary binders are -40.09 Kcal/mole and -41.00 Kcal/mole. The tertiary binders have -25.99 Kcal/mole and -29.32 Kcal/mole. Thus, we observe that the primary and secondary binders, through their large negative free energy changes provide a major structural stability to the 16S rRNA-protein complex in both the species.

However, the primary binders of *T. thermophilus* associate with 16S rRNA with higher BSA and higher ΔG compared to those of *E. coli*. This might entail the fact that at its high temperature habitat *T. thermophilus* 16S rRNA faces higher probability of unfolding/denaturation compared to that of *E. coli* at the earliest stages of 30S subunit assembly. Therefore, *T. thermophilus* requires proteins with higher ΔG of association to prevent the unfolding of 16S rRNA at the initial stages of 30S subunit protein-rRNA assembly, compared to those of *E. coli*. We do also observe that the BSAs of primary and secondary binders of the individual species are higher than those of tertiary binders (Table-1).

**3.3.2 Secondary and Tertiary binders form stable protein-protein interfaces:** In case of ribosomal protein-protein interactions, we observe that secondary and tertiary binders from highly stabilized interfaces as they interact mutually or with themselves. This general trend shows some variations in the individual species as well. An average secondary-tertiary interface is associated with -1.37 Kcal/mole free energy in *T. thermophilus* and -1.22 Kcal/mole free energy in *E. coli*. Primary-secondary interfaces have average -2.63 Kcal/mole free energy in *T. thermophilus* and -1.57 Kcal/mole free energy in *E. coli* (Supplementary Table S7). The high conformational changes of *T. thermophilus* proteins due to complexation with 16S rRNA follows from their high BSA and high ΔG of association compared to those of *E. coli*.

**3.3.3 Ribosomal protein-RNA interfaces compensate low base/backbone ratio by high interface polarity:** The double stranded A-form structure of the 16S rRNA causes most of its bases unavailable for proteins to form H-bonds, since they already have been used up for base pair formation. Therefore, a major selectivity is accepted by the ribosomal proteins to form H-bonds with the sugar-phosphate backbone of RNA. Jones *et al.* (2001) reported for non-ribosomal protein-RNA interactions that the interface H-bonding between proteins and RNA follows equal selection for RNA bases and backbone, thereby resulting base/backbone ratio to be unity. The base/backbone ratio of number of Hydrogen bonds formed between RNA and proteins gives us a demonstration of the major selection of RNA backbone for H-bond formation by the ribosomal proteins.

A comparison of the Hydrogen bond density of non-ribosomal (Jones *et al.*, 2001) and ribosomal protein-RNA interfaces (Table-4) shows that the later is higher than the former, with the special features of the ribosomal proteins having very low base/backbone ratio and high interface polarity. Non-ribosomal protein-RNA interface polarity was calculated to be 45.7% by Jones *et al.*. In addition, average ribosomal protein-RNA interface maintains 1.87 H-bonds per 100 $Å^2$ BSA, which is significantly higher than that of non-ribosomal (1.2 H-bonds per 100 $Å^2$ BSA). Having most of the H-bond donor and acceptor groups of rRNA bases being used up for base pair formation, the ribosomal protein-RNA interfaces face the availability of lesser number of H-bonding partners from nucleic acid bases (mainly backbone is available), compared to the non-ribosomal proteins (bases are available in higher propensity) per unit buried surface area. However, a high protein-rRNA interface polarity (Table S12) causes a relative increment of the surface density of available hydrogen bonding partners. Ribosome protein-rRNA interfaces have higher number of H-bonds among positive charged residues of proteins and negatively charged backbone of rRNA. Thus, an H-bonding surface density higher than the non-ribosomal protein-RNA interfaces is maintained in protein-rRNA interfaces even with a low base/backbone ratio.

The observed lower base/backbone ratio refers to the fact that the ribosomal H-bonding formation favors the backbone of RNA compared to the nucleotide bases and our result is consistent with a very recent observation (Gupta and Gribskov, 2011). Here, we extend their work by observing that *T. thermophilus* maintains a lower base/backbone ratio compared to *E. coli*. The *T. thermophilus* bacteria, being a habitat of high temperature environment, needs stronger protein-RNA contacts for gaining high stabilization and keep its ribosome structure intact, compared to *E. coli*. In addition to it, *T. thermophilus* faces a problem of low base/backbone ratio (0.75) compared to that of *E. coli* (0.83). Thus, for generating a high surface density of Hydrogen bonds at the protein-RNA interfaces, it has evolved with a high interface polarity.

**Table-4:** The comparison of *T. thermophilus* and *E. coli* and of non-ribosomal and ribosomal protein-RNA interface properties is shown in this table. See Supplementary section 1.2 and 1.3 for definitions of parameters. Surface density of H-bonding is expressed as number of H-bonds per 100 Å$^2$ BSA.

| Protein-RNA interface parameters | *T. thermophilus* values | *E. coli* values | Average values for protein-rRNA interfaces | Values for Non-ribosomal protein-RNA interfaces [33] |
|---|---|---|---|---|
| Base/Backbone ratio | 0.75 | 0.83 | 0.79 | 1.00 |
| Interface Polarity | 64.13% | 62.5% | 63.34% | 45.7% |
| Surface density of H-bonding | 1.94 | 1.79 | 1.87 | 1.20 |

**3.3.4  Polarity is conserved in evolutionary strategy of ribosomal proteins:** These pictures suggest that the interface polarity is adjusted in the evolutionary strategy of ribosome to adopt with the habitat conditions and keep structures intact. On the other hand, a comparison of ribosomal and non-ribosomal protein-RNA interfaces shows that due to structural constraints the former had to keep high interface polarity to assure highly stabilized protein-rRNA interfaces. High interface polarity can be maintained in two ways: one, by conserving interface residues (interface residues are mostly positively charged amino acids) and two, by replacing the interface residues by those of same charge. We have confirmed both these evolutionary strategy in Table S9a and Table S9b. The 'residue conservation factor' and 'charge conservation factor' values of ribosomal protein-RNA and protein-protein interfaces being greater than unity confirms our arguments (see materials and methods for definition of these two parameters). Thus, the necessity for maintenance of high polar interface (which seems to be a general stabilization strategy for protein-rRNA interfaces) also modulates the evolutionary trend of ribosomal proteins.

**3.3.5  Overall comparison between the stabilization strategy of *T. thermophilus* and *E. coli*:** From the previous results of individual species, we find an interesting general trend (with few exceptions). The proteins of *T. thermophilus* interact with 16S rRNA with higher BSA and higher ΔG than those of *E. coli* and consequently experience higher conformational changes. In addition, the protein-RNA interfaces of *T. thermophilus* are rich in hydrogen bonds and van der Waals contacts compared to those of the *E. coli*. This general trend can be explained by the fact that *E. coli* is habituated to live in normal conditions (room temperature) and *T. thermophilus* is habituated in extreme conditions (high temperature). *T. thermophilus*, even at high temperature, probably maintains the structural stability of its 30S ribosomal subunit through higher negative free energy of protein-rRNA association, H-bonding, and van der Waals contact; and the higher conformational change of the ribosomal proteins helps to attain these.

# 4   CONCLUSIONS

The dataset of 21 small ribosomal subunit structures is the largest dataset analyzed in this kind of analysis to date. Our work provides a number of new insights to the structure and stabilization of the 30S ribosomal complex. At the initial stages of the ribosomal association pathway, primary binder proteins provide high free energy of association and prevent its unfolding. Consequently, secondary binders join the subunit and form highly stabilized interfaces both with the 16S rRNA and primary binders; and generate potential site of interaction with the tertiary binder proteins. Tertiary binders finalize the assembled 30S structure and they mainly form stabilized protein-protein interfaces.

We do not only predict the conformational changes of the ribosomal proteins upon interaction with 16S rRNA, further linearly correlate them with the BSA and ΔG values of association. Conformational changes of proteins have been explained using their structural properties and binding strategy with 16S rRNA. We have identified the flexible and disordered proteins of 30S subunit. This linear relationship is confirmed in two bacterial and a eukaryotic species. We propose that this linearity might be a general rule of ribosomal protein-RNA assembly and will be followed in any ribosome of any species.

Further, we show that interface polarity is regulated in evolutionary strategy of ribosomal proteins. High polarity ensures high surface density of H-bonds and thereby provides structural stability to the 30S subunit. Therefore, the environmental conditions sweep up the interfaces of the ribosome to alter its physical states.

## ACKNOWLEDGEMENTS

The authors acknowledge the computational facility of DIC and Department of Biophysics, Molecular Biology and Bioinformatics, University of Calcutta. This research work is not funded by any authority; the authors performed it on their personal interests.

## SUPPLEMENTARY MATERIALS

Supplementary Materials and Methods (section 1), Tables S1-S12, Figures F1, Discussion (Section 2) and References (section-3) are available online.

# SUPPLEMENTARY MATERIALS

## 1. MATERIALS AND METHODS (in details)

### 1.1 DATASET

We collected a large set of 21 ribosomal complexes from PDB on 16/05/2012. Eleven of them are *Thermus thermophilus* (bacteria), nine from *Escherichia coli* (bacteria), and one from *Tetrahymena thermophila* (eukaryote). The PDB codes of these complexes with additional informations and citations are summarized in Table S1. We neglected all the structures containing modified residues, full length tRNA, mRNA, elongation factors and release factors. Except them, these 15 structures were collected from the whole PDB archive. Structures with ≥4.0 Å resolution are also disregarded. However, 1XNQ and 1IBK contains small fragments of mRNA (3-4 nucleotides only). Unique structures containing only ribosomal RNA and ribosomal proteins (there may be antibiotics and other small molecules bound) were selected for our analysis. Yeast ribosomal complexes are available in PDB, but as we will explain in the following, detectable protein-rRNA interfaces were not identified in them. Therefore, we used the only available *Tetrahymena thermophila* ribosomal structure, in spite of presence of modified residues in it.

**Table S1.** The Dataset of 21 ribosomal complexes collected from PDB.

| Organism | PDB-ID | Resolution (Å) | Citation |
|---|---|---|---|
| *Thermus thermophilus* | 1FJG | 3.00 | Carter, A.P., Clemons Jr., W.M., Brodersen, D.E., Morgan-Warren, R.J., Wimberly, B.T., Ramakrishnan, V. (2000) Nature **407**: 340-348 |
| | 1N36 | 3.65 | Ogle, J.M., Murphy IV, F.V., Tarry, M.J., Ramakrishnan, V. (2002), Cell(Cambridge,Mass.) **111**: 721-732 |
| | 1XNQ | 3.05 | Murphy, F.V., Ramakrishnan, V. (2004) Nat.Struct.Mol.Biol. **11**: 1251-1252 |
| | 2E5L | 3.30 | Kaminishi, T., Wilson, D.N., Takemoto, C., Harms, J.M., Kawazoe, M., Schluenzen, F., Hanawa-Suetsugu, K., Shirouzu, M., Fucini, P., Yokoyama, S. (2007) Structure **15**: 289-297 |
| | 2F4V | 3.80 | Murray, J.B., Meroueh, S.O., Russell, R.J., Lentzen, G., Haddad, J., Mobashery, S. (2006) Chem.Biol. **13**: 129-138 |
| | 1IBK | 3.31 | Ogle, J.M., Brodersen, D.E., Clemons Jr., W.M., Tarry, M.J., Carter, A.P., Ramakrishnan, V. (2001) Science **292**: 897-902 |
| | 2ZM6 | 3.30 | Kaminishi, T., Wang, H., Kawazoe, M., Ishii, R., Hanawa-Suetsugu, K., Nomura, M., Takemoto, C., Shirouzu, M., Paola, F., Yokoyama, S. (to be published) |
| | 3OGE | 3.00 | Bulkley, D., Innis, C.A., Blaha, G., Steitz, T.A. (2010). Revisiting the structures of several antibiotics bound to the bacterial ribosome. Proc.Natl.Acad.Sci.USA 107: 17158-17163 |
| | 3UXT | 3.20 | Bulkley, D., Johnson, F., Steitz, T.A., (2012) J.Mol.Biol. 416: 571-578 |
| | 3OHC | 3.00 | Bulkley, D., Innis, C.A., Blaha, G., Steitz, T.A., (2010) Proc.Natl.Acad.Sci.USA 107: 17158-17163 |
| | 2HHH | 3.35 | Schluenzen, F., Takemoto, C., Wilson, D.N., Kaminishi, T., Harms, J.M., Hanawa-Suetsugu, K., Szaflarski, W., Kawazoe, M., Shirouzo, M., Nierhaus, K.H., Yokoyama, S., Fucini, P., (2006) Nat.Struct.Mol.Biol. 13: 871-878 |
| *Escherichia coli* | 2AVY | 3.46 | Schuwirth, B.S., Borovinskaya, M.A., Hau, C.W., Zhang, W., Vila-Sanjurjo, A., Holton, J.M., Cate, J.H. (2005) Science **310**: 827-834 |
| | 3OAQ | 3.25 | Dunkle, J.A., Xiong, L., Mankin, A.S., Cate, J.H. (2010) Proc.Natl.Acad.Sci.USA **107**: 17152-17157 |
| | 2QAN | 3.21 | Borovinskaya, M.A., Pai, R.D., Zhang, W., Schuwirth, B.S., Holton, J.M., Hirokawa, G., Kaji, H., Kaji, A., Cate, J.H. (2007) Nat.Struct.Mol.Biol. **14**: 727-732 |
| | 3DF1 | 3.50 | Borovinskaya, M.A., Shoji, S., Fredrick, K., Cate, J.H.D. (2008) Rna **14**: 1590-1599 |
| | 2VHO | 3.00 | Bingel-Erlenmeyer, R., Kohler, R., Kramer, G., Sandikci, A., Antolic, S., Maier, T., Schaffitzel, C., Wiedmann, B., Bukau, B., Ban, N. (2008) Nature **452**: 108 |
| | 2QP0 | 3.50 | Borovinskaya, M.A., Shoji, S., Holton, J.M., Fredrick, K., Cate, J.H. (2007). A steric |
| | | | block in translation caused by the antibiotic spectinomycin. Acs Chem.Biol. 2: 545-552 |
| | 2QBB | 3.54 | Borovinskaya, M.A., Pai, R.D., Zhang, W., Schuwirth, B.S., Holton, J.M., Hirokawa, G., Kaji, H., Kaji, A., Cate, J.H., (2007) Nat.Struct.Mol.Biol. 14: 727-732 |
| | 2QOY | 3.50 | Borovinskaya, M.A., Shoji, S., Holton, J.M., Fredrick, K., Cate, J.H., (2007) Acs Chem.Biol. 2: 545-552 |
| | 1VS5 | 3.46 | Schuwirth, B.S., Day, J.M., Hau, C.W., Janssen, G.R., Dahlberg, A.E., Cate, J.H.D., Vila-Sanjurjo, A., (2006) Nat.Struct.Mol.Biol. 13: 879-886 |
| *Tetrahymena thermophila* | 2XZM | 3.93 | Rabl, J., Leibundgut, M., Ataide, S.F., Haag, A., Ban, N. (2011) Science **331:** 730-736 |

All the ribosomal proteins were analyzed in terms of sequence and three dimensional structural alignment, using PDBeFold server of EBI (http://www.ebi.ac.uk/msd-srv/ssm/cgi-bin/ssmserver). The root mean square deviation (RMSD) of two aligned protein structures in their three dimensional complexed conformation by SSM algorithm (RMSD of distances between Cα-atoms of matched residues at best 3D superposition of the template and target structures) is defined as:

$$RMSD = \sqrt{Ga^2 + Gb^2 - 2\frac{\sum_i \mathcal{R}ai \cdot \mathcal{R}bi}{\sqrt{\sum_i \mathcal{R}ai^2 \sum_i \mathcal{R}bi^2}} GaGb}$$

Where $Ga$ and $Gb$ are the radius of gyration for structures A and B, $\mathcal{R}ai$ and $\mathcal{R}bi$ are the co-ordinate vectors of Cα-atoms of proteins A and B after global superposition. Generally, the larger the RMSD, the distant the two matched structures are (Krissinel, 2004; Edger 2004). As suggested by Krissinel and Henrick in 2004 (Krissinel and Henrick, 2004), RMSD cannot be considered as a good measure of three dimensional structure dissimilarity when the target structure is a little distorted replica of the template (by a few Å), because the quality of alignment is often improved by unmapping the Cα atoms of less similar parts. Usually this is achieved by introducing a cut-off distance of about 2-4Å in SSM algorithm. Only that structure alignment is considered to be of a good quality that provides minimum RMSD for an alignment of maximum number of residues ($N_{align}$). Therefore, a quality factor has been suggested to quantify numerically the quality of alignment (denoted by Q) and it is defined as:

$$Q = N_{align}^2 / [\{1 + (RMSD/R_0)^2\} N_1 N_2]$$

Here $N_1$ and $N_2$ are the number of residues present in template and target structures. $R_0$ is an empirical parameter (chosen at 3Å) that measures the relative significance of RMSD and $N_{align}$ (Krissinel and Henrick, 2004). The Q-value is unity only for identical structures. The RMSD and Q-scores of 3D structural alignment were calculated for the multiple structural alignments of all corresponding ribosomal proteins for the 15 ribosomal complexes, using PDBeFold server. The amino acid sequence identity between the corresponding proteins was also analyzed quantitatively using the same server, where similarity between identical sequences is represented by unity. The more the identity factor is close to unity, the similar the two sequences are.

## 1.2 ANALYSIS OF PROTEIN-rRNA AND PROTEIN-PROTEIN INTERFACES

The structural characterization of protein-protein and protein-RNA interfaces were first conducted by Janin *et al.* in 1988, Miller in 1989, Argos in 1988 and Thornton and coworkers (Jones and Thornton, 1995, 1996; Jones et al., 1999, 2001), which concentrated both on the protein-protein and protein-nucleic acid interfaces characterized in terms of hydrophobicity, accessible surface area, interface shape and residue propensities. The comparisons of different types of protein complexes (enzyme-inhibitor and antibody-antigen) in terms of interface size and hydrophobicity have been analyzed by Janin and coworkers (Janin and Chothia, 1990; Duquerroy et al., 1991). Korn and Burnett in 1991, Young et al. in 1994 and Thornton et and coworkers have contributed progressively on the protein-protein interfaces thereafter. The researches on protein-nucleic acid interfaces started with some pioneer works of Steitz in 1990, Harrison in 1991 and Thornton and coworkers (Jones et al., 1999; Luscombe et al., 2000; Jones et al., 2001), where they concentrated on the characterization of the corresponding interfaces. Olson and co-workers (Olson 19996; Olson and Zhurkin, 1996; Olson et al., 1998) detected the DNA structures distort upon protein binding which was also supported by Dickerson in 1998. These works also have revealed the sequence dependent nature of the protein-DNA binding sites. However, the protein-RNA interfaces have also been analyzed in detail, with the example of Thornton et al. in 2001.

In case of protein-protein contacts, an amino acid residue is defined as an interface residue if it loses >1Å$^2$ of its accessible surface area when passing from uncomplexed state to complexed state (Jones et al., 1995). Jones et al. in 1999 and 2001 has shown that this is applicable to protein-RNA and protein-DNA complexes as well. By calculating the accessible surface area of the amino acid residues in the complexed three dimensional conformation both in complexed state (protein-RNA) and being isolated from the complex (only protein), it is possible to identify the amino acid residues with accessible surface area reduced >1Å$^2$ on complex formation with ribosomal RNA. They are known as the interface residues and the total number of interface residues in a single protein defines its RNA binding site (Janin et al., 1998). The Surface Racer program (Tsodikov et al., 2002) was used to calculate the accessible and buried surface areas of the proteins and RNA, with probe radius taken to be 1.4Å, which resembles the radius of one water molecule. We calculated the water accessible surface area (ASA) of the two interacting partners using Surface racer program separately (in their complexed conformation) and in associated state. If the ASA of the two partners are A1 and A2 and of their associated structure is A3, then buried surface area is

$$BSA = 1/2\,\{(A1 + A2) - A3\}$$

These results were compared with those provided by PDBePisa (Krissinel and Henrick, 2005, 2007; Krissinel 2009) web server of EBI (URL: http://ebi.ac.uk/msd-srv/prot-int/cgi-bin/piserver).

The %polarity of any protein-protein or protein-nucleic acid interface is defined as:

$$\%Polarity = \frac{BSA\,(polar)}{BSA\,(protein)} \times 100$$

where BSA (polar) is the buried surface area of polar atoms and BSA (protein) is the buried surface area of the total protein due to complex formation. For each ribosomal protein, this equation was used to calculate the %polarity of its interface with 16S rRNA. We also calculated the data of solvation free energy of association of each ribosomal protein that indicate the thermodynamic stability of every protein-protein and protein-RNA interface from PDBePisa server (Krissinel and Henrick, 2005, 2007; Krissinel 2009), compiled, and presented accordingly.

## 1.3  ANALYSIS OF ATOM-ATOM CONTACTS

We focused on the main two kinds of atom-atom contacts stabilizing the protein-rRNA interfaces: hydrogen bonds and van der Waals contacts. Intermolecular hydrogen bonds were calculated for protein-RNA complexes using NCONT program of ccp4i program suite (Winn et al., 2011), Discovery Studio Visualizer program (Accelrys Software Inc.) and UCSF Chimera software (Pettersen et al., 2004). The theoretical criteria used to define a hydrogen bond was the D-A distance ≤3.35Å, and D-H-A angle to be within $90^0$-$180^0$, where D stand for the donor group and A for the acceptor group. The van der Waals contacts between protein and RNA were defined as all contacts between atoms excluding those involved in hydrogen bonds that are ≤4.0Å apart (Jones et al., 2001). Distance criteria for van der Waals contact within protein are taken to be ≤5.0Å. The hydrogen bonding and van der Waals contact information was also calculated from the PDBePisa server (Krissinel and Henrick, 2005, 2007; Krissinel 2009) of EBI for all 15 ribosomal complexes, and were compared with Discovery Studio Visualizer and UCSF Chimera output. For each ribosomal protein, the surface density of Hydrogen bonding, defined as the number of hydrogen bonds per 100Å$^2$ buried surface area due to complex formation, was calculated from the available data, both at its RNA and protein-binding sites. All the data were compiled and presented according to the classifications adopted for the ribosomal proteins.

## 1.4  PROTEIN CONFORMATIONAL CHANGES DUE TO COMPLEX FORMATION

Binding of proteins with other proteins or nucleic acids is usually accompanied by significant changes of their three-dimensional conformations. There are prior works that predicted that there is a relationship between the conformational changes a protein undergoes upon complex formation and the interface size of the resulting complex. Janin et al. (4) predicted that complexes with large interfaces must require major structural changes upon complex formation due to the excessive accessible surface of their isolated subunits. Lo Conte et al. in 1999) showed that the subunits of protein complexes with large interfaces (BSA >2000Å$^2$) tend to undergo greater conformational changes upon complex formation, compared to those with smaller interfaces. The connection between the intrinsic flexibility of unbound proteins and the conformations they adopt upon binding has been analyzed in detail by Marsh and Teichmann in 2011. Analyzing accessible surface area of 4988 monomeric proteins Marsh and Teichmann were able to find out a relationship between molecular weight (M) and accessible surface area for all monomers:

$$ASA = 4.84 M^{0.760}$$

Since this equation is highly predictive of the ASA of monomeric and folded proteins, the authors concluded that the deviation of ASA from its predicted value might be a useful indicator of to what extent a protein resembles a folded monomer. They defined relative solvent accessible surface area, denoted by $ASA(rel)$ as observed ASA scaled by its predicted value from the molecular weight:

$$ASA(rel) = \frac{ASA(observed)}{ASA(predicted)}$$

The $ASA(rel)$ value for each subunit is to be considered in its bound conformation, isolated from the rest of the complex. By plotting the $ASA(rel)$ with RMSD values of complexed and uncomplexed conformations, the authors have observed the following relationships:

$$RMSD\ (homomer) = \exp[6.14 \times ASA(rel)^{complexed} - 5.95]$$

$$RMSD\ (heteromer) = \exp[6.35 \times ASA(rel)^{complexed} - 6.05]$$

Applying this algorithm proposed by Teichmann et al., we have tried to predict the three dimensional conformational changes of all the r-proteins due to their transition from the complexed state to the uncomplexed state. Since ribosomal proteins are all heteromers, the second equation will tell us the RMSD values between the complexed and uncomplexed states of each ribosomal protein. The ASA of the ribosomal proteins in their complexed conformation isolated from ribosomal assembly were calculated using Surface Racer program (22), and exact molecular weights were calculated using Discovery Studio Visualizer software (27). All the available data were compiled according to the classifications of r-proteins we adopted.

The Marsh and Teichmann algorithm we applied here describes a strategy to predict whether a protein structure is intrinsically disordered or not. The $ASA(rel)$ value of one monomer in bound state being ≥1.2 indicates a structurally flexible protein; whereas a value ≥ 1.4 indicates an intrinsically disordered protein structure.

## 1.5     ANALYSIS OF SEQUENCE CONSERVATION AT THE BINDING SITE

We are interested about the extent of amino acid sequence conservation of the binding sites of the ribosomal proteins, compared to that of the rest of protein surface. The conservation of the sequences at the binding sites since conservation of sequences or their substitution by residues of same chemical property is necessary for structural integrity. Taking under consideration 20 bacterial species from the bacterial phylogeny and of diverse availability in nature, we have tried to analyze the degree of conservation of the amino acid residues at the binding sites. The 20 species are *Escherichia coli, Vibrio cholerae, Staphylococcus aureus, Thermus thermophilus, Salmonella enterica, Micobacterium tuberculosis, Helicobacter pylori, Mycoplasma crocodyli, Actinobacter baumannii, Agrobacterium radiobacter, Cellulophaga lytica, Streptococcus pneumoniae, Persephonella marina, Nitrobacter hamburgensis, Pseudomonas putida, Rhodospirillum rubrum, Anabaena variabilis, Caulobacter crescentus, Chlorobium luteolum,* and *Actinomyces oris.* They contain a large variety of archaea, proteobacteria, firmicutes, actinobacteria, and modern bacteria, with observable diversity in availability in nature as well. The sequences of all the ribosomal proteins for these 20 bacterial species were collected from NCBI and using MUSCLE multiple sequence alignment program (Edgar, 2004) they were aligned together. The aligned sequences were then represented graphically as sequence logos using WebLogo server (Crooks et al., 2004) according to the probability of occurrence of them in the 20 different species. This gives us a visualized demonstration of the sequence similarity, conservation and the replacement of various amino acid residues throughout the evolution. In the WebLogo output, the sequences corresponding to *E. coli* that take part in the interactions (protein-protein and protein-RNA) were marked with information about both the van der Walls and Hydrogen bonding contacts at the binding sites.

To compute numerically the conservation of residues at the binding sites, we are proposing a new parameter: "the binding site residue conservation factor," denoted by $C_T$, which is defined as:

$$C_T = \frac{N_C^B / N^B}{N_C^S / N^S}$$

where, $N_C^B$ is the number of amino acid residues at the interfaces that are conserved to a cutoff value denoted by T, normally expressed in percentage of conservation (e.g. 70% cutoff conservation, T=70%). $N^B$ is the total number of interface residues; $N_C^S$ is the number of amino acid residues at the protein surface excluding binding sites that are conserved to a cutoff value denoted by T and $N^S$ is the total number of surface residues excluding interface. The residue conservation factor being >1, we can say that the corresponding protein shows higher tendency in preserving the amino acid sequences at the binding sites, compared to the rest of its surface residues. Varying the cutoff value of conservation from 50% to 100% at a step of 10%, we have calculated $C_T$ in each case.

The binding site of proteins does not only contain all the conserved residues, but there are many residues (identified for *E. coli*) that have been replaced by other residues in other bacterial species. We have observed that the substitutions of the residues at the binding sites follow a distinct pattern: there is a slight higher tendency of substituting the residues by those of similar charge (positive-to-positive, negative-to-negative, and neutral-to-neutral) at any given position compared to rest parts of the proteins (discussed in detail in Results and Discussions). To verify this, we are proposing another factor, the "interface charge conservation factor," denoted by $P_C$ as:

$$P_c = \frac{N_C^B / N_T^B}{N_C^S / N_T^S}$$

where $N_C^B$ is the number of replacements of the amino acid residues by those of the same charge (positive-to-positive, negative-to-negative, neutral-to-neutral), plus the number of conserved residues at the interface. $N_T^B$ is the total number of replacements of all residues at the interface, $N_C^S$ is the number of the residues replaced by those of the same charge, plus the number of conserved residues at the protein surface excluding interface and $N_T^S$ is the number of replacement of all residues at the surface excluding interface. At any given position at the binding site, if one residue is conserved, or is substituted by other residues of same charge, then we can say that the residue positional charge is conserve there. The interface residue conservation factor computationally measures the higher tendency of conserving the residue positional charge at the interface, compared to rest of the protein surface, if its numerical value is >1. In this algorithm, it is considered that one amino acid has been replaced by others of same charge if the probability of occurrence of all residues of the same charge at that position is ≥90%, or is conserved if its probability of occurrence at that position is ≥90%.

## 1.6 STATISTICAL TESTS AND CALCULATIONS

### 1.6.1 DATA CLUSTERING

The protein-RNA and protein-protein interface properties were clustered together to find out the high importance of a group of proteins over the others in ribosomal stabilization. Simple Hierarchical clustering algorithms (Gronau and Moran, 2007) were used in our analysis to identify groups of proteins with some unique interface features. Globally closest pair clustering algorithms was tested on our data samples and simple UPGMA (Unweighted Pair Grouping Method with Arithmetic mean) is selected for all the analysis. In this method, the input is a dissimilarity matrix *D* over a set of elements *S*.

<u>Initialization</u>: A cluster set *C* is initialized by defining a singleton cluster $C_i = \{i\}$ for every element $i \in S$. Output hierarchy is initialized $H \Leftarrow C$.
<u>Loop:</u> While $|C| > 1$, the following steps are accepted:

1. Cluster pair selection: A pair of distinct clusters is selected $\{C_i, C_j\} \subseteq C$ of minimum dissimilarity under *D*.

2. Cluster pair joining: $C_i$, $C_j$ are removed from the cluster set $C$ and is replaced by $C_i \cup C_j$. $C_i \cup C_j$ cluster is then added to the hierarchy $H$.

3. The dissimilarity $D\left(C_k, (C_i \cup C_j)\right)$ for every $C_k \in C' \setminus (C_i \cup C_j)$ is calculated.

<u>Finalize</u>: Return to the hierarchy $H$.

The dissimilarity matrix in UPGMA method defines the dissimilarity between clusters as their average dissimilarity. This is achieved by using the following reduction formula (ref. 36):

$$D\left(C_k, (C_i \cup C_j)\right) \leftarrow \frac{|C_i|}{|C_i|+|C_j|} D(C_k, C_i) + \frac{|C_j|}{|C_i|+|C_j|} D(C_k, C_i)$$

The commutativity property holds trivially when the reduction formula $F$ induces a dissimilarity function $D_F$ over the set of all clusters, s.t. $D_F(C_1, C_2)$ depends only on the dissimilarities between elements in $C_1 \cup C_2$. Such are for instance the reduction formulae for UPGMA (ref. 32):

$$D_{UPGMA}(C, C') = \frac{1}{|C||C'|} \sum_{i \in C, j \in C'} D(i, j)$$

So, utilizing UPGMA clustering method, groups of proteins sharing common features are identified. This enables us to identify the roles of different groups of proteins in structural assembly.

### 1.6.2 MANN-WHITNEY U-TESTS

Uniqueness (in terms of structural features) of each of the protein groups were statistically tested in terms of Mann-Whitney U-test for comparison (Mann and Whitney, 1947). The null hypothesis adopted: the two groups of data selected for analysis originate from the same population. The alternate hypothesis is: the two sets of data selected for analysis originate from two different populations. The null hypothesis is rejected and the alternate hypothesis is accepted when $p \leq 0.01$. When $0.01 < p < 0.05$, we concluded that the two sets of data are marginally different (different populations). The two groups of data are considered to be originated from the same population, if $p \geq 0.05$.

All the clustering algorithms and U-tests were carried on using PAST statistical software package (Hammer et al., 2001). In this work, we also have tried to correlate some interface parameters, the correlation coefficients of whom were also calculated using PAST. The statistics of various linear and surface fitting were calculated using Origin data analysis and graphic workspace (OriginLab Corporation). Origin was also used to produce graph plots included in this supplementary material.

## 2. RESULTS AND DISCUSSION

### 2.1 RESULTS

**Table S2:** The classifications of the ribosomal proteins of *E. coli* and *T. thermophilus* according to their association with the 16S RNA and according to their individual influence in structure stabilization.

| Ribosomal protein classifications according to their association with the complex | | Ribosomal protein classifications according to their influence in 30S structure stabilization | |
|---|---|---|---|
| Classifications | Proteins | Classifications | Proteins |
| Primary Binders | S4, S7, S8, S15, S17 & S20 | Group-1 | S4, S9, S12 |
| Secondary Binders | S9, S12, S13, S16, S18 & S19 | Group-2 | S16, S17, S20 |
| Tertiary Binders | S2, S3, S5, S6, S10, S11, S14, S21 & THX | Group-3 | S3, S5, S7, S8, S10, S11, S13, S14, S15, S19 |
| | | Group-4 | S2, S6, S18, S21, THX |

**Table S3:** The buried surface area of the ribosomal proteins for *Thermus thermophilus* and *Escherichia coli*.

| Ribosomal proteins | Average BSA of *T. thermophilus* proteins (Å²) | Standard Deviation | BSA of *E. coli* proteins (Å²) | Standard deviation |
|---|---|---|---|---|
| S2 | 998.71 | 74.43 | 937.12 | 88.90 |
| S3 | 1691.35 | 56.04 | 1622.40 | 86.32 |
| S4 | 2965.99 | 243.30 | 3006.92 | 93.88 |
| S5 | 1959.26 | 53.15 | 1769.82 | 54.93 |
| S6 | 647.18 | 189.11 | 432.66 | 30.73 |
| S7 | 1859.82 | 76.37 | 1790.46 | 84.70 |
| S8 | 1874.48 | 77.83 | 1742.46 | 24.97 |
| S9 | 2943.62 | 173.86 | 2836.02 | 43.67 |
| S10 | 1801.79 | 40.92 | 1742.60 | 27.61 |
| S11 | 1825.82 | 31.41 | 2000.72 | 6.52 |
| S12 | 3106.87 | 118.09 | 2751.24 | 151.00 |
| S13 | 2155.44 | 128.88 | 1808.47 | 49.70 |
| S14 | 1931.28 | 99.15 | 2072.64 | 101.79 |
| S15 | 1776.25 | 36.98 | 1628.04 | 36.62 |
| S16 | 2505.85 | 69.35 | 2099.58 | 53.48 |
| S17 | 2669.90 | 72.11 | 1603.21 | 37.27 |
| S18 | 864.57 | 14.37 | 935.93 | 40.09 |
| S19 | 1549.82 | 162.29 | 1550.89 | 91.95 |
| S20 | 2541.10 | 88.14 | 2139.38 | 47.71 |
| S21 | - | - | 514.84 | 36.25 |
| THX | 1287.49 | 34.63 | - | - |

**Table S4:** The free energy of association of the ribosomal proteins with 16S rRNA for *Thermus thermophilus* and *Escherichia coli*.

| Ribosomal proteins | Solvent free energy of association (ΔG) with 16S rRNA (Kcal/mole) for *T. thermophilus* proteins | Standard Deviation | Solvent free energy of association (ΔG) with 16S rRNA (Kcal/mole) for *E. coli* proteins (Å²) | Standard deviation |
|---|---|---|---|---|
| S2 | -17.58 | 4.17 | -19.28 | 2.68 |
| S3 | -28.49 | 1.79 | -32.68 | 0.85 |
| S4 | -52.60 | 2.36 | -61.79 | 3.75 |
| S5 | -33.96 | 1.56 | -30.32 | 0.97 |
| S6 | -8.18 | 0.30 | -8.10 | 0.62 |
| S7 | -25.05 | 3.47 | -26.26 | 6.02 |
| S8 | -28.46 | 0.75 | -28.06 | 1.93 |
| S9 | -54.36 | 2.20 | -51.54 | 4.52 |
| S10 | -28.37 | 3.05 | -26.65 | 2.14 |
| S11 | -32.58 | 0.84 | -34.22 | 1.45 |
| S12 | -58.13 | 1.40 | -50.99 | 1.85 |
| S13 | -42.98 | 4.91 | -42.38 | 0.75 |
| S14 | -31.68 | 3.57 | -41.24 | 2.61 |
| S15 | -37.01 | 1.67 | -29.98 | 2.27 |
| S16 | -45.88 | 4.38 | -37.11 | 2.35 |
| S17 | -42.38 | 2.03 | -33.09 | 1.36 |
| S18 | -20.17 | 0.63 | -19.02 | 1.46 |
| S19 | -19.01 | 3.26 | -19.10 | 3.58 |
| S20 | -51.56 | 1.52 | -44.66 | 1.64 |
| S21 | - | - | -9.80 | 2.06 |
| THX | -27.09 | 3.17 | - | - |

**Table S5:** Mann-Whitney U-test results. In this table, the average values of the physical parameters between the two bacterial species (*T. thermophilus* and *E. coli*) are used for each protein of the groups.

| Interface Properties | U-test between groups | p-value | Two groups are from same/different populations |
|---|---|---|---|
| Buried Surface Area | Group-1 vs. Group-2 | 0.0994 | Same population |
| | Group-2 vs. Group-3 | 0.00699 | Different populations |
| | Group-3 vs. Group-4 | 0.000666 | Different populations |
| | Group-4 vs. Group-1 | 0.03572 | Different populations, difference is marginally significant |
| | Group-1 vs. Group-3 | 0.006993 | Different populations |
| | Group-2 vs. Group-4 | 0.03571 | Different populations, difference is marginally significant |
| | Primary vs. Secondary | 0.9048 | Same populations |
| | Secondary vs. tertiary | 0.2571 | Same populations |
| | Tertiary vs. Primary | 0.04262 | Different populations, difference is marginally significant |
| | Primary and secondary vs. tertiary | 0.03142 | Different populations, difference is marginally significant |
| Free energy of association with 16S rRNA | Group-1 vs. group-2 | 0.1 | Same population |
| | Group-2 vs. Group-3 | 0.02797 | Different populations, difference is marginally significant |
| | Group-3 vs. Group-4 | 0.002664 | Different populations |
| | Group-4 vs. Group-1 | 0.03571 | Different populations, difference is marginally significant |
| | Group-1 vs. group-3 | 0.006993 | Different populations |
| | Group-2 vs. Group-4 | 0.03571 | Different populations, difference is marginally significant |
| | Primary vs. Secondary | 0.7302 | Same populations |
| | Secondary vs. tertiary | 0.1714 | Same populations |
| | Tertiary vs. Primary | 0.08125 | Same populations |
| | Primary and secondary vs. tertiary | 0.04734 | Different populations, difference is marginally significant |

**Table S6:** The free energy of association, Surface density of Hydrogen bond and van der Waals contacts at protein-rRNA interface of ribosomal proteins are displayed. The proteins are classified into four distinct groups according to importance in ribosomal structure stabilization. From this table, it is clear that Group-1 proteins are most important is stabilization of ribosomal complex with highest free energy of association and highest surface density of atomic contacts. Group-4 proteins are least important.

| Protein Groups | Solvent free energy of association (ΔG) with 16S rRNA (Kcal/mole) | Average BSA of ribosomal proteins at rRNA interfaces (Å$^2$) |
|---|---|---|

| Group-1 | -54.90 (2.14) | 2935.11 (48.60) |
| Group-2 | -42.45 (5.25) | 2259.84 (108.40) |
| Group-3 | -30.92 (6.42) | 1807.69 (143.05) |
| Group-4 | -15.77 (7.73) | 842.08 (322.49) |

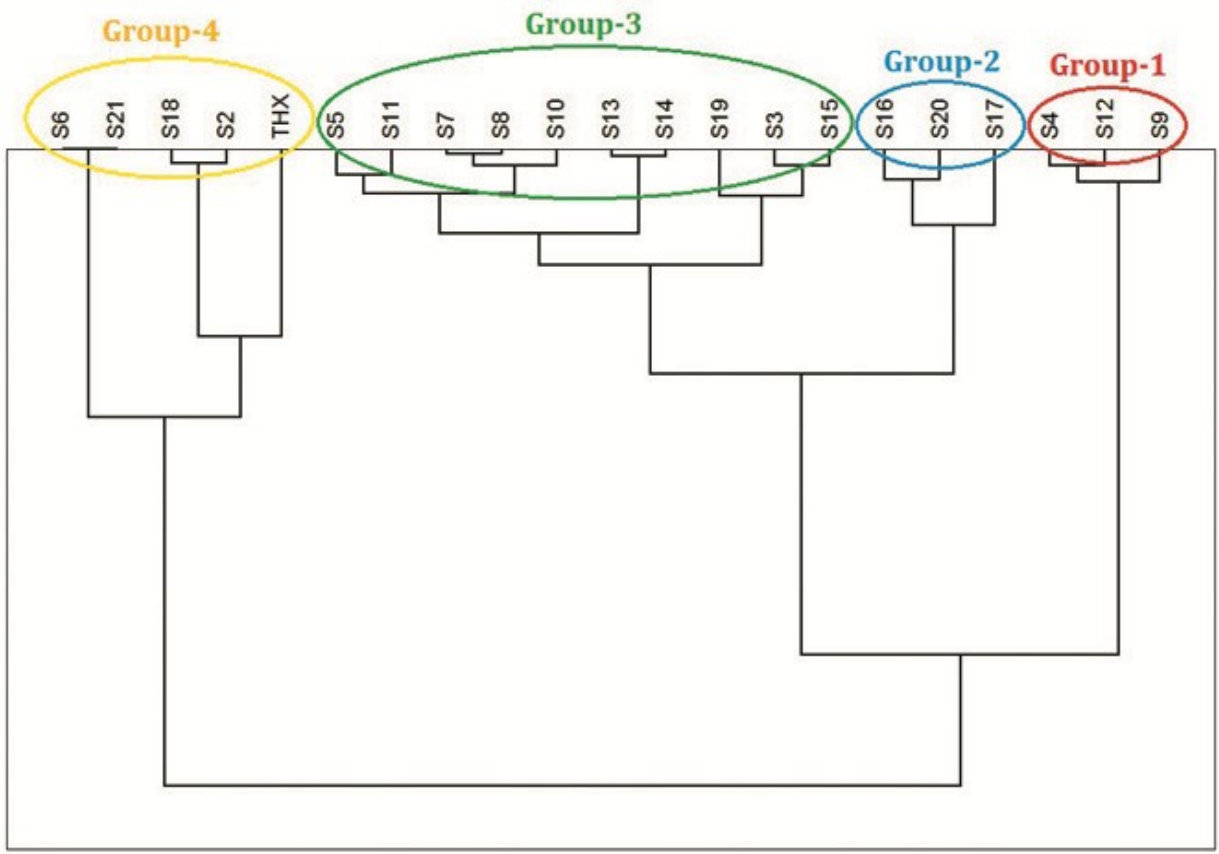

**Figure F1:** The dendogram of the BSA-based clustering of the small-subunit ribosomal proteins using UPGMA method. Proteins of the Group-1 and Group-2 have the largest interface areas with 16S rRNA. Proteins of Group-4 have minimum interface areas. Proteins of Group-3 have moderately large interface areas. The image is produced by using PAST statistical software package (34).

**Table S7:** The protein-protein interface properties of the 30S ribosomal subunit. We can see that the available protein-protein interfaces are different for *T. thermophilus* and *E. coli*. From this table, we also observe that some of the smallest protein-protein interfaces do not contain Hydrogen bonds. For example, S2-S5 interface in *E. coli* do not contain any Hydrogen bonding.

| Protein-protein interfaces | Buried surface area at the interface (Å²) | | Free energy of interface formation (Kcal/mole) | | Surface Density of Hydrogen bonds (No. of bonds per 100 Å² BSA) | | Surface Density of van der Waals contacts (No. of contacts per 100 Å² BSA) | |
|---|---|---|---|---|---|---|---|---|
| | T. thermophilus | E. coli | T. thermophilus | E. coli | T. thermophilus | E. coli | T. thermophilus | E. coli |
| S2-S3 | 56.3 | - | 1.2 | - | 2.1 | - | 6.3 | - |
| S2-S5 | 65.6 | 29.5 | 1.1 | 0.7 | - | - | 6.1 | 3.6 |
| S2-S8 | 258.7 | 2.9 | -0.9 | 0.1 | 1.3 | - | 13.4 | |
| S3-S4 | 37.4 | 20.0 | 0.8 | 0.6 | 5.2 | - | 11.6 | 3.8 |
| S3-S5 | 77.2 | 145.2 | 0.7 | 1.1 | 1.3 | 1.3 | 6.3 | 5.8 |
| S3-S10 | 396.1 | 326.5 | -3.5 | -4.7 | 0.5 | 0.5 | 10.9 | 6.7 |
| S3-S14 | 831.6 | 883.7 | -8.8 | -9.1 | 0.6 | 0.9 | 10.4 | 13.4 |
| S4-S5 | 354.0 | 264.2 | 0.6 | -0.8 | 0.8 | 0.8 | 5.9 | 11.3 |
| S5-S8 | 566.9 | 487.2 | -1.6 | -4.0 | 1.0 | 0.7 | 11.7 | 12.5 |
| S6-S18 | 1027.5 | 669.9 | -8.2 | -6.9 | 35.7 | | 15.6 | |
| S7-S9 | 251.7 | 92.6 | -3.6 | -0.8 | 0.2 | 0.2 | 6.0 | 0.7 |
| S7-S11 | 234.3 | 182.9 | -1.3 | -1.9 | 0.6 | 3.3 | 9.2 | 19.3 |
| S8-S12 | 57.9 | 58.0 | -1.0 | -0.7 | 0.0 | | 4.7 | 0.8 |
| S8-S17 | 125.2 | 122.0 | -0.5 | -0.9 | 1.7 | 3.5 | 8.2 | 18.1 |
| S9-S10 | 160.0 | 137.2 | -0.6 | -0.9 | 0.8 | 0.8 | 9.1 | 4.3 |
| S9-S13 | 68.1 | 76.9 | 0.1 | 0.4 | 0.9 | 0.7 | 11.1 | 10.0 |
| S9-S14 | 809.4 | 763.8 | -8.2 | -8.0 | 2.0 | | 6.3 | |
| S10-S14 | 311.4 | 122.6 | -1.0 | 0.3 | 8.9 | 2.1 | 14.3 | 10.1 |
| S11-S18 | 207.2 | 196.1 | -3.3 | -3.2 | 0.6 | 0.1 | 12.1 | 2.5 |
| S12-S17 | 402.9 | 364.3 | -3.5 | -2.0 | 0.3 | 2.0 | 5.8 | 30.9 |
| S13-S19 | - | 988.9 | - | -6.0 | 1.0 | 7.8 | 6.6 | 80.8 |
| S11-S21 | - | 248.0 | - | -2.3 | - | 0.5 | - | 3.5 |

**Table S8a:** The $A_{REL}$ values (expressing the structural flexibility of proteins) of small subunit ribosomal proteins are enlisted here. Data are calculated according to Marsh and Teichmann algorithm (section 1.4). The $A_{REL}>1.2$ indicated structural flexibility and $A_{REL}>1.4$ indicates intrinsic disorder of corresponding proteins. Averaging between two species, we see S14 is intrinsically disordered, and S12, S13 has a very high structural flexibility. On the other hand, S2, S6, S16 are structurally very rigid proteins.

| Ribosomal proteins | $A_{REL}$ values for T. thermophilus | Standard Deviation | $A_{REL}$ values for E. coli | Standard deviation | Average $A_{REL}$ |
|---|---|---|---|---|---|
| S2 | 1.10 | 0.04 | 1.12 | 0.05 | 1.11 |
| S3 | 1.20 | 0.03 | 1.22 | 0.05 | 1.22 |
| S4 | 1.16 | 0.02 | 1.19 | 0.05 | 1.17 |
| S5 | 1.12 | 0.03 | 1.12 | 0.03 | 1.12 |
| S6 | 1.03 | 0.03 | 1.16 | 0.05 | 1.10 |
| S7 | 1.22 | 0.05 | 1.27 | 0.04 | 1.25 |
| S8 | 1.10 | 0.02 | 1.12 | 0.04 | 1.11 |
| S9 | 1.21 | 0.02 | 1.24 | 0.05 | 1.23 |
| S10 | 1.27 | 0.02 | 1.29 | 0.05 | 1.28 |
| S11 | 1.20 | 0.02 | 1.22 | 0.03 | 1.21 |
| S12 | 1.33 | 0.03 | 1.41 | 0.05 | 1.37 |
| S13 | 1.28 | 0.03 | 1.36 | 0.06 | 1.32 |
| S14 | 1.40 | 0.03 | 1.41 | 0.04 | 1.41 |
| S15 | 1.17 | 0.01 | 1.22 | 0.06 | 1.20 |
| S16 | 1.13 | 0.01 | 1.08 | 0.04 | 1.11 |
| S17 | 1.21 | 0.02 | 1.30 | 0.05 | 1.26 |
| S18 | 1.14 | 0.03 | 1.20 | 0.04 | 1.17 |
| S19 | 1.20 | 0.02 | 1.25 | 0.05 | 1.23 |
| S20 | 1.23 | 0.01 | 1.23 | 0.22 | 1.23 |
| S21 | - | - | 1.18 | 0.04 | 1.18 |
| THX | 1.49 | 0.06 | - | - | 1.49 |

**Table S8b:** The constant terms (a, b and c) and the surface fitting statistics of three parameters of ribosomal proteins: Free energy of association, Buried surface area and RMSD between complexed and uncomplexed states. Predicted equation is:
$$\Delta G = a + b(BSA) + c(RMSD)$$

| Constant terms | *Thermus thermophilus* | *Escherichia coli* | *Tetrahymena thermophila* |
|---|---|---|---|
| a | 2.099 | 4.421 | -2.259 |
| b | -0.009 | -0.231 | -0.018 |
| c | -0.019 | -0.019 | 0.257 |
| R-value of surface fitting | 0.95 | 0.94 | 0.93 |
| R-square value of surface fitting | 0.89 | 0.89 | 0.86 |

**Table S9a:** The residue conservation factor and charge conservation factor for ribosomal proteins classified according to association at different conservation cutoffs.

| Protein Classifications | Residue Conservation factor according to protein classification (at various conservation cutoffs) | | | | | | Charge Conservation factor |
|---|---|---|---|---|---|---|---|
| | 50% conservation | 60% conservation | 70% conservation | 80% conservation | 90% conservation | 100% conservation | |
| Primary | 1.20 | 1.32 | 1.48 | 1.56 | 1.65 | 2.01 | 1.36 |
| Secondary | 1.26 | 1.37 | 1.44 | 1.51 | 1.76 | 2.07 | 1.30 |
| Tertiary | 1.30 | 1.42 | 1.45 | 1.57 | 2.05 | 3.79 | 1.32 |

**Table S9b:** The residue conservation factor for ribosomal proteins at different conservation cutoffs.

| Ribosomal Proteins | Residue conservation factor (various cutoffs) | | | | | | Charge conservation factor |
|---|---|---|---|---|---|---|---|
| | 50% conservation | 60% conservation | 70% conservation | 80% conservation | 90% conservation | 100% conservation | |
| S2 | 1.71 | 2.05 | 2.01 | 2.67 | 3.14 | 4.47 | 2.49 |
| S3 | 1.14 | 1.25 | 1.38 | 1.44 | 1.43 | 2.44 | 1.57 |
| S4 | 1.08 | 1.24 | 1.22 | 1.41 | 1.63 | 2.49 | 1.52 |
| S5 | 1.29 | 1.31 | 1.64 | 1.75 | 1.77 | 3.62 | 1.14 |
| S6 | 1.71 | 1.66 | 1.97 | 1.71 | 3.41 | 8.53 | 1.39 |
| S7 | 0.99 | 1.04 | 1.07 | 1.04 | 0.85 | 0.55 | 0.92 |
| S8 | 1.38 | 1.38 | 1.42 | 1.91 | 2.79 | 3.12 | 1.40 |
| S9 | 1.48 | 1.84 | 1.97 | 1.85 | 1.89 | 1.86 | 1.63 |
| S10 | 1.41 | 1.61 | 1.46 | 1.67 | 1.75 | 3.38 | 1.17 |
| S11 | 1.10 | 1.32 | 1.40 | 1.64 | 1.65 | 1.72 | 1.03 |
| S12 | 1.07 | 1.04 | 1.04 | 1.02 | 1.11 | 1.15 | 1.01 |
| S13 | 1.61 | 1.75 | 1.97 | 2.18 | 2.24 | 2.56 | 1.84 |
| S14 | 0.99 | 1.15 | 1.11 | 1.11 | 1.21 | 2.38 | 0.97 |
| S15 | 1.49 | 1.61 | 1.99 | 1.83 | 1.96 | 2.74 | 1.77 |
| S16 | 1.05 | 1.12 | 1.24 | 1.60 | 2.42 | 2.23 | 1.07 |
| S17 | 1.08 | 1.06 | 1.21 | 1.09 | 1.02 | 1.13 | 1.09 |
| S18 | 1.29 | 1.24 | 1.17 | 1.08 | 1.36 | 3.09 | 0.89 |
| S19 | 1.06 | 1.22 | 1.26 | 1.33 | 1.53 | 1.53 | 1.38 |
| S20 | 1.17 | 1.61 | 1.97 | 2.07 | - | - | 1.46 |
| S21 | 1.04 | 1.04 | 0.63 | 0.56 | - | - | 0.81 |

Large proteins like S3, S4, S5 and S6 also show high binding site residue conservation factor. However, there is almost no correlation identified between the conservation tendency and protein size, association or architecture.

**Table S10:** The surface density of Hydrogen bonds at the protein-rRNA interfaces of *Thermus thermophilus* and *Escherichia coli* ribosomes.

| Ribosomal proteins | Surface Density of H-bonds ($\sigma(H)$) at protein-rRNA interface (No. of bonds per 100Å² BSA) for *T. thermophilus* proteins | Standard Deviation | Surface Density of H-bonds ($\sigma(H)$) at protein-rRNA interface (No. of bonds per 100Å² BSA) for *E. coli* proteins (Å²) | Standard deviation |
|---|---|---|---|---|
| S2 | 1.24 | 0.41 | 1.31 | 0.48 |
| S3 | 1.74 | 0.34 | 1.72 | 0.41 |
| S4 | 2.32 | 0.38 | 1.80 | 0.26 |
| S5 | 2.09 | 0.24 | 1.78 | 0.42 |
| S6 | 2.17 | 0.44 | 1.26 | 0.50 |
| S7 | 2.37 | 0.53 | 1.74 | 0.38 |
| S8 | 1.81 | 0.28 | 1.98 | 0.16 |
| S9 | 2.28 | 0.61 | 2.20 | 0.29 |
| S10 | 1.44 | 0.34 | 1.63 | 0.20 |
| S11 | 1.93 | 0.29 | 1.80 | 0.26 |
| S12 | 2.21 | 0.33 | 2.17 | 0.27 |
| S13 | 2.20 | 0.40 | 2.21 | 0.34 |
| S14 | 1.71 | 0.44 | 2.34 | 0.31 |
| S15 | 1.65 | 0.30 | 1.67 | 0.35 |
| S16 | 1.73 | 0.18 | 1.67 | 0.15 |
| S17 | 1.95 | 0.29 | 1.74 | 0.25 |
| S18 | 2.00 | 0.28 | 2.39 | 0.28 |
| S19 | 1.91 | 0.26 | 1.92 | 0.39 |
| S20 | 1.98 | 0.36 | 1.68 | 0.39 |
| S21 | - | - | 0.82 | 0.33 |
| THX | 2.14 | 0.36 | - | - |

**Table S11:** The surface density of van der Waals contacts at the protein-rRNA interfaces of *Thermus thermophilus* and *Escherichia coli* ribosomes.

| Ribosomal proteins | Surface Density of van der Waals contacts ($\sigma(v)$) at protein-rRNA interface (No. of bonds per 100Å² BSA) for *T. thermophilus* proteins | Standard Deviation | Surface Density of van der Waals contacts ($\sigma(v)$) at protein-rRNA interface (No. of bonds per 100Å² BSA) for *E. coli* proteins (Å²) | Standard deviation |
|---|---|---|---|---|
| S2 | 12.6 | 1.2 | 10.9 | 1.8 |
| S3 | 16.8 | 0.9 | 16.8 | 1.2 |
| S4 | 15.3 | 1.3 | 14.7 | 0.5 |
| S5 | 14.4 | 0.8 | 15.6 | 1.2 |
| S6 | 14.4 | 1.5 | 9.0 | 2.2 |
| S7 | 19.1 | 1.4 | 15.7 | 1.4 |
| S8 | 14.9 | 0.9 | 17.0 | 1.2 |
| S9 | 18.7 | 1.2 | 16.7 | 1.2 |
| S10 | 15.9 | 0.8 | 14.6 | 0.8 |
| S11 | 17.4 | 1.5 | 17.3 | 0.6 |
| S12 | 15.8 | 0.8 | 16.1 | 0.8 |
| S13 | 19.4 | 1.9 | 18.9 | 2.1 |
| S14 | 15.2 | 1.9 | 16.0 | 0.2 |
| S15 | 15.6 | 1.0 | 15.3 | 1.4 |
| S16 | 13.6 | 0.7 | 12.7 | 0.4 |
| S17 | 16.6 | 1.7 | 13.8 | 1.4 |
| S18 | 13.8 | 0.8 | 16.2 | 2.5 |
| S19 | 19.7 | 1.0 | 19.7 | 1.6 |
| S20 | 16.1 | 0.8 | 13.8 | 0.7 |
| S21 | - | - | 10.3 | 2.9 |
| THX | 13.7 | 0.8 | - | - |

**Table S12:** The Base/Backbone ratio and the %polarity data of the ribosomal proteins for *T. thermophilus* and *E. coli*.

| Ribosomal Proteins | Interface Polarity | | Base/backbone ratio | |
|---|---|---|---|---|
| | *T. thermophilus* | *E. coli* | *T. thermophilus* | *E. coli* |
| S2 | 65.12 | 62.04 | 0.97 | 0.69 |
| S3 | 61.53 | 60.31 | 0.91 | 0.81 |
| S4 | 68.30 | 62.59 | 0.39 | 0.49 |
| S5 | 63.32 | 60.01 | 0.73 | 0.96 |
| S6 | 62.70 | 59.95 | 0.34 | 0.37 |
| S7 | 68.78 | 57.30 | 0.74 | 1.32 |
| S8 | 54.88 | 59.09 | 1.77 | 2.27 |
| S9 | 66.45 | 59.77 | 0.35 | 0.40 |
| S10 | 59.15 | 59.07 | 0.73 | 0.79 |
| S11 | 64.37 | 62.27 | 0.80 | 1.08 |
| S12 | 68.39 | 70.04 | 0.42 | 0.59 |
| S13 | 65.23 | 65.19 | 0.43 | 0.15 |
| S14 | 67.38 | 64.30 | 0.46 | 0.46 |
| S15 | 61.58 | 64.17 | 1.07 | 2.02 |
| S16 | 62.59 | 62.04 | 0.57 | 0.50 |
| S17 | 62.92 | 61.37 | 0.81 | 0.85 |
| S18 | 63.20 | 61.90 | 0.81 | 0.87 |
| S19 | 59.72 | 63.99 | 1.82 | 0.84 |
| S20 | 68.99 | 68.69 | 0.78 | 0.35 |
| THX | - | 66.98 | - | 0.75 |
| S21 | 67.97 | - | 0.13 | |